\documentclass[prb,aps,floatfix,twocolumn,amsmath,amssymb,showpacs,superscriptaddress]{revtex4-1}
\newcommand{\ket}[1]{\ensuremath{\left|{#1}\right\rangle}}
\newcommand{\bra}[1]{\ensuremath{\left\langle{#1}\right|}}

\usepackage{epsfig}
\usepackage{hyperref}
\newcommand{\be}{\begin{equation}}
\newcommand{\ee}{\end{equation}}
\def\nn{\nonumber\\}
\def\eps{\epsilon}

\begin{document}
\title{Low-Temperature Dynamical Structure Factor of the Two-Leg
  Spin-\texorpdfstring{$\frac12$}{1/2} Heisenberg Ladder}
\date{24 July 2010}
\author{W. D. Goetze}
\affiliation{Rudolf Peierls Centre for Theoretical Physics, University of Oxford, Oxford, OX1 3NP, UK}
\author{U. Karahasanovic}
\affiliation{School of Physics and Astronomy, North Haugh, St Andrews, Fife, KY16 9SS, UK}
\author{F. H. L. Essler}
\affiliation{Rudolf Peierls Centre for Theoretical Physics, University of Oxford, Oxford, OX1 3NP, UK}
\pacs{75.10.Jm, 75.10.Pq, 75.40.Gb}
\begin{abstract}
We determine the dynamical structure factor of the two-leg spin-$\frac12$
Heisenberg ladder at low temperatures in the regime of strong rung
coupling. The dominant feature at zero temperature is the coherent
triplon mode. We show that the lineshape of this mode broadens in a 
non-symmetric way at finite temperatures and that the degree of
asymmetry increases with temperature. We also show that at low
frequencies a temperature induced resonance akin to the Villain mode
in the spin-$\frac12$ Heisenberg Ising chain emerges.
\end{abstract}

\maketitle
\section{Introduction}
\begin{figure}[bht]
\includegraphics[width=0.3\textwidth]{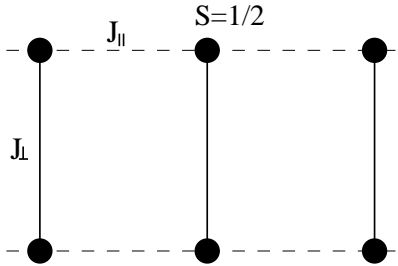}
\caption{Exchange couplings for a spin-ladder system. In the strong
rung coupling limit $J_{\parallel } \ll J_{\perp }$.}  
\label{ladder}
\end{figure}

The zero temperature behaviour of two-leg spin-$\frac12$ Heisenberg ladders
is by now well understood and has been analyzed by a variety of
theoretical methods \cite{hida,DRS,GRS,BR,SNT,OSW,HS,kotov,KSU,SU}. 
Recently, the dynamical structure factor (DSF) has been measured by
inelastic neutron scattering experiments for the ladder compounds $\rm
La_4Sr_{10}Cu_{24}O_{41}$ \cite{notbohm} and ${\rm CaCu_2O_3}$ \cite{bella}
and was found to be in excellent agreement with theoretical predictions at
$T=0$. The limit of strong rung coupling $\alpha=J_\parallel/J_\perp \ll 1$, see
Fig.~\ref{ladder}, is particularly simple. In the limit $\alpha=0$, the
ground state is a tensor product state of rung singlets. Excitations
involve breaking one of the dimers, which leads to a finite gap
$\Delta=J_\perp$. A small but finite $J_\parallel$ gives a dispersion
to these excitations, which are commonly referred to as either
``magnons'' or ``triplons'' \cite{SU}. We will follow the latter
terminology in this work. The triplon bandwidth is small
compared to their gap.   
The dominant feature of the DSF at zero temperature is a
delta-function following the triplon dispersion.  At higher energies
there are multi-triplon continua, which for small $\alpha$ are
weak. These features have been analyzed in detail in the literature
\cite{hida,DRS,GRS,BR,SNT,OSW,kotov,KSU,SU,HS}.

Much less is known about the finite temperature dynamics 
of one-dimensional quantum magnets in general and the two-leg ladder
in particular
\cite{damle,RMK,DS05,doyon1,doyon2,RZ,RT,AKT,DG,Mikeska,EK08,JEK08}.
In the limit of large $\alpha$, the DSF
for the two-leg ladder model was studied by means of a semiclassical
analysis by Damle and Sachdev \cite{damle}. They showed that at very 
low temperatures $T\ll\Delta$ the $T=0$ triplon delta-function peak in
the DSF broadens and is well described by a Lorentzian lineshape. This
behaviour was argued to be universal for one-dimensional gapped
antiferromagnets. Very recently the question of how the DSF evolves as
the temperature is increased above the semiclassical regime has been
addressed in several models by numerical \cite{Mikeska} and analytical
methods \cite{EK08,JEK08}. It was shown that at higher 
temperatures, but still smaller than the gap, the triplon peak is
broadened in a rather asymmetric fashion.
In this paper we calculate the DSF
for a spin-ladder system (Fig.~\ref{ladder}) at low
temperatures. This is a quantity of experimental interest, probed by
inelastic neutron scattering experiments
\cite{exp1,exp2,exp3,exp4,exp5,exp5a,exp6,exp7}. Our calculation
is restricted to the limit of weak coupling between the dimers, which
we treat in perturbation theory to first order in $\alpha =
J_{\parallel}/J_{\perp }$ for both excitation energies and matrix elements.

The Hamiltonian of the spin-ladder system reads
\begin{align}
\label{e1}
\mathcal{H}&=\mathcal{H}_0+\mathcal{H}_1, \nn
\mathcal{H}_1&= \sum_{j=0}^1 \sum_{n=0} ^{L-1} J_{\parallel } \mathbf
        {S}_{j,n} \cdot \mathbf {S}_{j,n+1 }, \nn
\mathcal{H}_0&=\sum_{n=0} ^{L-1} J_{\bot } \mathbf{S}_{0,n} \cdot \mathbf{S}_{1,n }.
\end{align}
Here the dominant exchange coupling~$J_{\bot}$ is along the rungs
connecting neighbouring spins on different legs of the ladder and 
$J_{\parallel} \ll J_{\bot}$ represents a small interaction between
the neighbouring rungs. In the limit of zero interrung coupling, the ground
state is a product of singlet states on every rung. The
elementary excitations are $S=1$ triplets of energy~$J_\perp$,
which is the difference between the dimer triplet and singlet states.

Our first goal is to calculate the dynamical susceptibility, which is related
to the DSF by
\begin{equation}
\label{e2}
S^{\alpha \gamma}(\omega, \mathbf{Q})=-\frac{1}{\pi}\frac{1}{1-\exp(-\beta \omega)}\Im\left[\chi^{\alpha \gamma}(\omega, \mathbf{Q})\right].
\end{equation}
Here $\alpha,\gamma=x,y,z$. In the Matsubara formalism, the $\alpha
\gamma$~component of the susceptibility is given by 
\begin{multline}
\label{e3}
\chi^{\alpha \gamma}(\omega, \mathbf{Q}) = -\frac{1}{2L}\int_{0}^{\beta} d\tau e^{i \omega_n \tau}  \\
\times \sum_{j,k=0}^{1}  \sum_{l,l'=0}^{L-1} e^{-i\mathbf{Q} \cdot (\mathbf{R}_{j,l}-\mathbf{R}_{k,l'})}\langle S^{\alpha}_{j,l}(\tau)S^{\gamma}_{k,l'} \rangle \biggr|_{\omega_n\rightarrow \eta-i\omega},
\end{multline}
where $\langle\ldots\rangle$ denotes the thermal average
\begin{equation}
\label{e4}
\langle \mathbf{S}^{\alpha}_{j,l}(\tau)\mathbf{S}^{\gamma}_{k,l'}  \rangle= \frac{1}{Z}\mathbf{Tr} \left (e^{-\beta \mathcal{H}}S^{\alpha}_{j,l}(\tau)S^{\gamma}_{k,l'}\right).  
\end{equation}
As a consequence of the $SU(2)$ symmetry of the Heisenberg interaction,
all off-diagonal elements of the susceptibility tensor are zero
and all diagonal elements are identical. It is therefore sufficient
to calculate~$\chi^{zz}(\omega, \mathbf{Q})$. The trace in \eqref{e4} is taken
over a basis of states, and $Z$ represents the partition
function. Using translational invariance, writing the time evolution
of spin operators as $S^z(\tau)$, and inserting a complete set of
simultaneous eigenstates of the Hamiltonian and the momentum operator
into the formula for the susceptibility \eqref{e3} gives
\begin{multline}
 \label{es1}
\chi^{zz}(\omega, \mathbf{Q})=- \frac{1}{Z} \int_{0}^{\beta} d \tau e^{i \omega_n \tau}\frac{1}{2L}  \sum_{l,l'} e^{-iQ_{\parallel} (l-l')} \\
\times \sum_{n,m} e^{-\beta \epsilon_{m}}  e^{-\tau(\epsilon_{n}-\epsilon_{m})} e^{i(p_{n}-p_{m})(l-l')} M_{n,m}.
\end{multline}
The sum runs over a complete set of states~$\ket{n}$ with well
defined momentum~$p_n$ and energy~$\epsilon_n$.  The expression for
$M_{n,m}$ is  
\begin{align}
\label{es2}
M_{n,m}=&\left|\bra{n}S^{z}_{0,0}\ket{m}\right|^2+\left|\bra{n}S^{z}_{1,0}\ket{m}\right|^2\nn
&+e^{iQ_\perp}\bra{n}S^{z}_{0,0}\ket{m}\bra{m}S^{z}_{1,0}\ket{n}\nn
&+ e^{-iQ_\perp}\bra{n}S^{z}_{1,0}\ket{m}\bra{m}S^{z}_{0,0}\ket{n}.
\end{align}

After performing the Fourier transform and analytically continuing to
real frequencies, Equation \eqref{es1} reads
\begin{equation}
\label{es3}
\chi^{zz}(\omega,\mathbf{Q})=\frac{L}{2}\sum_{n,m} \frac{e^{-\beta \epsilon_{n}}-e^{-\beta \epsilon_{m}}}{\omega+i\eta+\epsilon_{n}-\epsilon_{m}} \delta_{Q_{\parallel}+p_{n},p_{m}}M_{n,m}.
\end{equation}

\section{Diagonalization of Short Chains}
\begin{figure}[tbp]
\includegraphics*[width=0.45\textwidth]{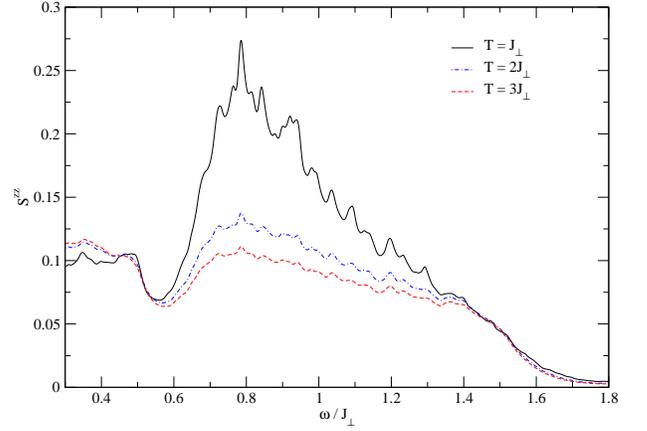}
\caption{The interband transition for $(Q_\parallel,Q_\perp)=(\pi,\pi/2)$ found by exact diagonalization of a $J_\parallel=0.25J_\perp$ ladder system of $L=8$ rungs. The thermal broadening is much greater than $\eta=0.01$.}  
\label{edhight}
\end{figure}

For small systems, we may calculate a basis of simultaneous
eigenstates of the Hamiltonian and the momentum operator numerically
using a standard exact diagonalization (ED) package.
This allows the spectral sum in Equation~\eqref{es3} to be evaluated.
As a ladder of $L$~rungs has a Hilbert space of dimension~$4^L$,
this method is only feasible up to
$L=8$. The numerically calculated DSF for such
small finite systems is obtained as a sum over delta functions in
frequency. In order to facilitate comparisons with the result in the
thermodynamic limit, we introduce a sufficiently large value
for the Lorentzian width~$\eta$ in \eqref{es3} to obtain a smooth
function. To observe thermal broadening of the lineshape, the
temperature has to be large enough for thermal effects to dominate
over the artificial broadening due to $\eta$. 
In Fig.~\ref{edhight}, we show some typical results obtained by this
method at intermediate temperatures~$T \agt J_\perp$. In
section~\ref{sec:conc}, we compare the results of the low-temperature expansion
developed in the following to the exact numerical answer for $L=8$.

\section{Low Temperature Expansion}
\label{sec:lowTex}
In what follows, we use the fact that for $J_\parallel\ll J_\perp$
states can still be labelled by their triplon number for
$J_\parallel=0$, although it ceases to be a good quantum number for
$J_\parallel\neq 0$. Subsequently, we will refer to the
perturbative eigenstates as ``$r$-particle states''~$\ket{\gamma_r}$,
where the terminology indicates that they reduce to $r$-triplon states
when $J_\parallel$ is taken to zero. Here, $\gamma_r$ is a complete set
of quantum numbers uniquely identifying the state under consideration. 
Using this notation, we rewrite Equation~\eqref{es3} as
\begin{align}
\label{esx3}
&\chi^{zz}(\omega,\mathbf{Q})\equiv\frac1Z \sum_{r,s=0}^\infty E_{r,s}+F_{r,s},\nn
&E_{r,s}=\frac{L}{2}\sum_{\gamma_r,\gamma_s} \frac{e^{-\beta
    \epsilon_{\gamma_r}}}{\omega+i\eta+\epsilon_{\gamma_r}-\epsilon_{\gamma_s}}
\delta_{Q_{\parallel}+p_{\gamma_r},p_{\gamma_s}}M_{\gamma_r,\gamma_s},\nn
&F_{r,s}=-\frac{L}{2}\sum_{\gamma_r,\gamma_s} \frac{e^{-\beta \epsilon_{\gamma_s}}}{\omega+i\eta+\epsilon_{\gamma_r}-\epsilon_{\gamma_s}} \delta_{Q_{\parallel}+p_{\gamma_r},p_{\gamma_s}}M_{\gamma_r,\gamma_s}.
\end{align}
For sufficiently small $J_\parallel\ll J_\perp$, we may associate a formal
temperature dependence with $E_{r,s}$ and~$F_{r,s}$
\be
E_{r,s}=\mathcal{O}\Bigl(e^{-\beta r J_\perp}\Bigr),
F_{r,s}=\mathcal{O}\Bigl(e^{-\beta s J_\perp}\Bigr).
\ee
Equation~\eqref{es2} becomes
\be
M_{\gamma_r,\gamma_s}=2\left|\bra{\gamma_r}S^{z}_{0,0}\ket{\gamma_s}\right|^2\left(1+(-1)^{r-s}\cos(Q_\perp)\right),
\ee 
because due to the leg exchange symmetry
\be
\bra{\gamma_r}S^{z}_{0,0}(0)\ket{\gamma_s} = (-1)^{r-s} \bra{\gamma_r}S^{z}_{1,0}(0)\ket{\gamma_s}.
\ee
The quantities $E_{r,s}$ and~$F_{r,s}$ as well as the partition
function~$Z$ diverge in the thermodynamic limit. We therefore reorder the
spectral sum in the spirit of a linked-cluster expansion
following Ref.~[\onlinecite{EK09}]. To do so, we express the partition
function as 
\be
Z=\sum_{n=0}^\infty Z_n,
\ee
where $Z_n$ is the contribution of $n$-particle states. It is
furthermore convenient to combine quantities with the same formal
temperature dependence as 
\be
G_{r,s}=E_{r,s}+F_{s,r}.
\ee
We then define the quantities
\begin{align}
\mathcal{C}_0=&\sum_{j=1}^\infty G_{0,j},\nn
\mathcal{C}_1=&G_{1,0}+\sum_{j=1}^\infty \Big(G_{1,j}-Z_1
G_{0,j-1}\Big),\nn
\mathcal{C}_2=&G_{2,0}+\Big(G_{2,1}-Z_1G_{1,0}\Big)\nn
&+\sum_{m=2}^\infty\Big(G_{2,m}-Z_1G_{1,m-1}+(Z_1^2-Z_2)G_{0,m-2}\Big)
,\nn
\mathcal{C}_3=&\dots
\end{align}
The $\mathcal{C}_n$ are the sums of all cluster functions with
the same formal temperature dependence. Hence we obtain by construction
that (as the triplon bandwidth is small compared to the triplon gap)
\be
\mathcal{C}_n=\mathcal{O}(e^{-\beta n J_\perp}).
\ee
We can then re-express the spectral sum in (\ref{es1}) as
\be
\chi^{zz}(\omega,{\bf Q})=\frac{1}{Z}\sum_{r,s=0}^\infty \left(E_{r,s}+F_{r,s}\right)
=\sum_{n=0}^\infty \mathcal{C}_n.
\label{lowTex}
\ee
Now we postulate that $\mathcal{C}_n$ are finite in the thermodynamic
limit and (\ref{lowTex}) constitutes a low-temperature expansion. This
assumption is valid in the limit of non-interacting dimers
$J_\parallel=0$. We furthermore verify it by explicit 
calculation for the leading contribution $\mathcal{C}_1$ for
$J_\parallel\neq 0$. 
By virtue of the existence of a spectral gap $\Delta$ the contribution
of $\mathcal{C}_n$ is seen to be proportional to $e^{-n \Delta/T}$, so that
(\ref{lowTex}) constitutes a low-temperature expansion in the small
parameter $e^{-\Delta/T}$, which can be viewed as the density of
triplons in the state of thermal equilibrium. 

\subsection{Divergences}
\label{sec:div}
As we will see, the expansion~\eqref{lowTex} exhibits ``infrared''
divergences at
\begin{enumerate}
\item{}
$\omega\to\pm\eps(Q_\parallel)$.
These occur in the ``interband transition'' terms $G_{j,j+1}$.
\item{}
$\omega\to \pm 2J_\parallel\sin(Q_\parallel/2) $.
These occur in the ``intraband transition'' terms $G_{j,j}$.
\end{enumerate}
In order to deal with these divergences, we need to sum up an infinite
number of terms in the low-temperature expansion. This can be done
by following Refs.~[\onlinecite{EK08,JEK08}].
\subsubsection{``Interband'' processes}
\label{sec:div1}
The expansion~\eqref{lowTex} contains as the leading term the $T=0$
result, which diverges when the external frequency $\omega$ approaches
the single-triplon dispersion $\eps(Q_\parallel)$ like
\be
\frac{1}{(\omega+i\eta)^2-\eps^2(Q_\parallel)}.
\ee
This corresponds to the coherent propagation of a single triplon at
$T=0$ and leads to a contribution proportional to
$\delta(\omega^2-\eps^2(Q_\parallel))$ in the DSF.
On the other hand, for any finite temperature we expect this
delta-function to be broadened. This is a non-perturbative effect and
cannot be captured in any finite order in the expansion
(\ref{lowTex}). The fact that a broadening occurs emerges from the
occurrence of ``infrared'' divergences in (\ref{lowTex}), i.e.\ 
singularities when the external frequency~$\omega$ approaches the
single-triplon dispersion~$\eps(Q_\parallel)$. For example, we show
below that the first sub-leading contribution~$\mathcal{C}_1$ exhibits a
divergence 
\be
\left(\frac{1}{(\omega+i\eta)^2-\eps^2(Q_\parallel)}\right)^2.
\ee
We expect the higher terms in the expansion to exhibit ever
stronger divergences of this type, which 
need to be summed up in order to obtain a physically meaningful
result. This can be achieved by employing a
self-energy formalism \cite{EK08,JEK08}. To deal specifically with the
divergence at $\omega^2=\eps^2(Q_\parallel)$, we divide the
expansion~\eqref{lowTex} for the susceptibility into a singular (for
$\omega^2\to\eps^2(Q_\parallel)$) and a
regular piece as follows 
\be
 \chi^{zz}(\omega, \mathbf{Q})=
 \chi^{zz}_{\mathrm{sing},1}(\omega, \mathbf{Q})+
 \chi^{zz}_{\mathrm{reg}}(\omega, \mathbf{Q}).
\label{chi_split1}
\ee
We then introduce a self-energy
$\Sigma_1(\omega, \mathbf{Q})$ by expressing the singular contribution to
the dynamical susceptibility in the form of
\begin{align}
\label{e6}
\chi^{zz}_{\mathrm{sing},1}(\omega, \mathbf{Q})
&=\frac{G_{0,1}(\omega,\mathbf{Q})}{1-G_{0,1}(\omega, \mathbf{Q})\Sigma_1(\omega,\mathbf{Q})} \nn  
& = G_{0,1}(\omega, \mathbf{Q})+G_{0,1}^2(\omega, \mathbf{Q})\Sigma_1(\omega, \mathbf{Q})+\cdots.
\end{align}
Here $G_{0,1}(\omega, \mathbf{Q})$ is the singular contribution to the
leading term~$\mathcal{C}_0$ in the expansion~\eqref{lowTex}. Matching
\eqref{e6} to \eqref{lowTex} then yields a low-temperature expansion
of both $\chi_{\mathrm{reg}}(\omega,{\bf Q})$ and the self-energy
\be
\Sigma_1(\omega,\mathbf{Q})=\sum_{j=1}^\infty\Sigma_1^{(j)}(\omega,\mathbf{Q}),
\ee
where the formal temperature dependence of the $n^{\mathrm{th}}$
contribution is
\be
\Sigma_1^{(n)}(\omega,\mathbf{Q})=\mathcal{O}\Bigl(e^{-n\beta\Delta}\Bigr).
\ee
\subsubsection{``Intraband'' Processes}
\label{sec:div2}
In the intraband processes~$G_{j,j}(\omega,\mathbf{Q})$ ($j=1,2$), we encounter
singularities of the form
\be
\Big[4J_\parallel^2\sin^2(Q_\parallel/2)-(\omega+i\eta)^2\Big]^{-j-1/2}.
\label{divintra}
\ee
We can deal with these singularities by employing a self-energy
formalism in a way completely analogous to the way we proceeded for
the interband processes. This results in a two-self-energy formalism
for the dynamical susceptibility. We express $\chi^{zz}(\omega,
\mathbf{Q})$ as a sum of three terms
\be
 \chi^{zz}(\omega, \mathbf{Q})=
 \chi^{zz}_{\mathrm{sing},1}(\omega, \mathbf{Q})+
 \chi^{zz}_{\mathrm{sing},2}(\omega, \mathbf{Q})+
 \chi^{zz}_{\mathrm{reg}}(\omega, \mathbf{Q}),
\ee 
where $\chi^{zz}_{\mathrm{sing},1}(\omega, \mathbf{Q})$ and
$\chi^{zz}_{\mathrm{sing},2}(\omega, \mathbf{Q})$ denote the
contributions of all terms singular for
$\omega^2\to\eps^2(Q_\parallel)$ and $\omega^2\to 
4J^2_\parallel\sin^2(Q_\parallel/2) $ respectively.
The contribution $\chi^{zz}_{\mathrm{sing},2}(\omega, \mathbf{Q})$ defines
a self-energy~$\Sigma_2(\omega, \mathbf{Q})$ by
\begin{align}
\label{singintra}
\chi^{zz}_{\mathrm{sing},2}(\omega, \mathbf{Q})&=\frac{G_{1,1}(\omega,
   \mathbf{Q})}{1-G_{1,1}(\omega, \mathbf{Q})\Sigma_2(\omega,
   \mathbf{Q})} \nn  
& =  G_{1,1}(\omega, \mathbf{Q})+G_{1,1}^2(\omega, \mathbf{Q})\Sigma_2(\omega, \mathbf{Q})+\dots\nn
\end{align}
Matching the expansions~\eqref{singintra} to the low-temperature
expansion for $\chi^{zz}_{\mathrm{sing},2}(\omega, \mathbf{Q})$ generates a
low-temperature expansion for the self-energy $\Sigma_2(\omega,
\mathbf{Q})$. 
\section{Excited states in the limit of weak interdimer coupling}
\subsection{Single triplon excited states}
We start with the Hamiltonian~\eqref{e1}. $\mathcal{H}_0$ is the dominant part
of the Hamiltonian, which describes $L$ uncoupled dimers.
The eigenstates of $\mathcal{H}_0$ are tensor products of singlet and triplet
states at sites $n=0,\ldots,L-1$. The unique ground state of $\mathcal{H}_0$ is thus
a series of singlet states on every site $n$. There are $3L$ degenerate
first excited states that consist of $L-1$ singlets and one triplet.
We treat $\mathcal{H}_1$ as a perturbation to $\mathcal{H}_0$ and construct a basis for
one- and two-particle excited states.

We define an operator~$d_a(m)$, which creates a triplet at site~$a$
with $z$-component of spin~$m$ when acting on the ground state~$\ket{0}$.
Single particle states with a definite value of momentum~$p$ that
carry spin-$1$ are constructed as
 \begin{equation}
 \label{e7}
 \ket{p,m} =\frac{1}{\sqrt{L}} \sum_{n=0} ^{L-1} e^{ipn}d_n(m)\ket{0}.
 \end{equation}
With periodic boundary conditions $S_L \equiv S_0$, translational
invariance makes momentum a good quantum
number and the above states are orthogonal, which enables us
to use non-degenerate perturbation theory to calculate the
single particle energy shifts. To first order in $\alpha=J_{\parallel}/J_{\perp}$,
the dispersion is given by 
\begin{equation}
\label{e8}
\epsilon_{p}=J_{\bot}+J_{\parallel}\cos(pa_\parallel),
\end{equation}
where $a_\parallel$ is the separation between the dimers.
Imposing periodic boundary conditions leads to the quantization of
one-particle momenta
\be
e^{ipL}=1.
\ee

\subsection{Two-triplon excited states}
\label{secsol2}
We now construct a basis of two particle states in which $\mathcal{H}_1$ is diagonal.
To lowest order in $\alpha$, the two-particle states can be written as
as 
 \begin{equation}
 \label{e9}
 \ket{p_1,p_2,S,m} =\mathcal{N}_S(p_1,p_2) \sum_{a=1} ^{L-1} \sum_{b=0} ^{a-1} \psi ^S_{a,b}(p_1,p_2) \phi^{S,m}_{a,b}\ket{0},
 \end{equation}
where
 \begin{equation}
 \label{e10}
 \phi^{S,m}_{a,b}=\sum_{m_1,m_2} \Phi^{S,m}_{m_1,m_2} d_a(m_1)d_b(m_2).
 \end{equation}
Here $\Phi^{S,m}$ are Clebsch-Gordan coefficients.
The total spin takes values $S=0,1,2$ and the normalization
$\mathcal{N}_S(p_1,p_2)$ depends on spin and linear momenta in general. 
 The spatial part of the wavefunction is given by
 \begin{equation}
 \label{e11}
 \psi^S_{a,b}(p_1,p_2)=e^{i(p_1a+p_2b)}+A^S_{p_1,p_2}e^{i(p_1b+p_2a)},
 \end{equation}
where the phase-shifts $A^S_{p_1,p_2}$ encode triplon-triplon interactions.
The boundary condition $\psi^S_{L-1,b}(p_1,p_2)\equiv (-1)^S \psi^S_{b,0}(p_1,p_2)$, where the
sign is due to odd $S$ states being antisymmetric, leads
to non-trivial quantization of two-particle momenta
 \begin{equation}
 \label{e12}
 (-1)^S A^S_{p_1,p_2}=e^{ip_1L}=e^{-ip_2L}.
 \end{equation}
 These equations require a numerical solution.
 Since for real momenta $A^S_{p_1,p_2}$ is a pure phase, we
 introduce the notation
 \begin{equation}
  \label{e13}
 \delta^S_{p_1,p_2}=-i\ln\left(A^S_{p_1,p_2}\right).
 \end{equation}
 The normalization of two-particle states is given by 
\begin{multline}
\label{e14}
\mathcal{N}_S(p_1,p_2)= \left[ L\left( L-1 \right)\right.\\
-\left.L\frac{\sin\left(\frac12(p_1-p_2)-\delta_{p_1p_2}^S\right)}{\sin\left(\frac12(p_1-p_2)\right)} \right]^{-1/2}.
\end{multline}
The two-particle states have degeneracy $3^2\binom{L}{2}$. A basis of
the two-particle subspace in which $\mathcal{H}_1$ is diagonal is
constructed by requiring that 
\begin{equation}
 \label{e15}
\mathcal{P}_2\mathcal{H}_1 \ket{p_1,p_2,S,m}=(\epsilon_{p_{1}}+\epsilon_{p_2})\ket{p_1,p_2,S,m}.
\end{equation}
Here $\mathcal{P}_2$ is the projection operator onto two-particle states. This
leads to a condition on $A^S_{p_1,p_2}$. When the triplets in the sum
(\ref{e8}) are not on adjacent rungs, this condition is satisfied
for any $A$. Considering the case of neighbouring triplets, we find 
\begin{align}
 \label{e16}
A^0_{p_1,p_2}&=-\frac{1+e^{-i(p_1+p_2)}+2e^{-ip_2}}{1+e^{-i(p_1+p_2)}+2e^{-ip_1}} \nn
A^1_{p_1,p_2}&=-\frac{1+e^{-i(p_1+p_2)}+e^{-ip_2}}{1+e^{-i(p_1+p_2)}+e^{-ip_1}} \nn
A^2_{p_1,p_2}&=-\frac{1+e^{-i(p_1+p_2)}-e^{-ip_2}}{1+e^{-i(p_1+p_2)}-e^{-ip_1}} .
\end{align}
The procedure for solving Equation~\eqref{e16} is outlined in Appendix~\ref{sec:bae}. Without affecting the result in the thermodynamic limit, we simplify the calculation by considering $L$ to be even.

\section{Matrix elements}
\begin{table*}[tb]
\caption{Non-zero matrix elements of $S^z_{j,0}$ to order $\alpha$. For the definitions see \eqref{e17}, \eqref{eq:intra} and \eqref{e21}.}
\label{tab1}
\begin{center}
\begin{tabular}{|| r @{$S^{z}_{j,0}$} l || l ||}
\hline
$\bra{0}$ & $\ket{p,0}$ & $(-1)^{j+1}\frac{1}{2\sqrt{L}}\left(1-\frac\alpha2 \cos(p)\right)$\\ [1ex]
$\bra{p',\pm 1}$ & $\ket{p,\pm 1}$ & $\pm\frac{1}{2L}$\\[1ex]
$\bra{p_1,p_2,0,0}$ & $\ket{p,0}$ & $(-1)^{j+1}\sqrt{\frac1{12L^3}}\left(U_{0}(p,p_1,p_2)-\frac\alpha2V_0(p,p_1,p_2)\right)$ \\[1ex]
$\bra{p_1,p_2,1,\pm1}$ & $\ket{p,\pm1}$ &$\pm(-1)^{j}\sqrt{\frac{1}{8L^3}}\left(U_{1}(p,p_1,p_2) - \frac\alpha2 V_1(p,p_1,p_2)\right)$\\[1ex]
$\bra{p_1,p_2,2,0}$ & $\ket{p,0}$ & $(-1)^{j}\sqrt{\frac{1}{6L^3}}\left(U_{2}(p,p_1,p_2)-\frac\alpha2 V_2(p,p_1,p_2)\right)$ \\[1ex]
$\bra{p_1,p_2,2,\pm1}$ & $\ket{p,\pm1}$ &$ (-1)^{j}\sqrt{\frac{1}{8L^3}}\left(U_{2}(p,p_1,p_2) - \frac\alpha2 V_2(p,p_1,p_2)\right)$\\[1ex]
$\bra{p^\prime_1,p^\prime_2,1,\pm1}$ & $\ket{p_1,p_2,1,\pm1}$ &$ \mp\frac{1}{4L^2}W_{1,1}(p^\prime_1,p^\prime_2,p_1,p_2) $\\[1ex]
$\bra{p^\prime_1,p^\prime_2,2,\pm2}$ & $\ket{p_1,p_2,2,\pm2}$ &$ \mp\frac{1}{2L^2}W_{2,2}(p^\prime_1,p^\prime_2,p_1,p_2) $\\[1ex]
$\bra{p^\prime_1,p^\prime_2,2,\pm1}$ & $\ket{p_1,p_2,2,\pm1}$ &$ \pm\frac{1}{4L^2}W_{2,2}(p^\prime_1,p^\prime_2,p_1,p_2) $\\[1ex]
$\bra{p^\prime_1,p^\prime_2,2,\pm1}$ & $\ket{p_1,p_2,1,\pm1}$ &$ \pm\frac{1}{4L^2}W_{2,1}(p^\prime_1,p^\prime_2,p_1,p_2) $\\[1ex]
$\bra{p^\prime_1,p^\prime_2,2,0}$ & $\ket{p_1,p_2,1,0}$ &$ \frac{1}{2\sqrt{3}L^2}W_{2,1}(p^\prime_1,p^\prime_2,p_1,p_2) $\\[1ex]
$\bra{p^\prime_1,p^\prime_2,1,0}$ & $\ket{p_1,p_2,0,0}$ &$ \frac{1}{2L^2}\sqrt{\frac23}W_{1,0}(p^\prime_1,p^\prime_2,p_1,p_2) $\\[1ex]
\hline
\end{tabular}
\end{center}
\end{table*}
\subsection{Selection rules}
At low temperatures $T\ll J_\perp$, the leading terms in the expansion~\eqref{lowTex}
involve states with at most two triplons
(in the aforementioned sense that the corresponding states reduce to
states with at most two triplets in the $J_\parallel=0$ limit). In the
following we compute the matrix elements which link $0,1$ and
$2$-particle states.

The operator $S^z_{j,l}$ acts on a single site, thus changing the triplon number by $\Delta n=0$ or~$1$. To first order in $\alpha$, $\mathcal{H}_1$ mixes states with those differing in triplon number by $\Delta n =\pm2$. As however we will only consider the modulus squared of the matrix elements, this correction is only relevant in the case that the matrix element is non-zero to leading order. The rule remains valid.
$S^z_{j,l}$ conserves the total $S^z$ which we have used to label states, so $\Delta S^z=0$.
The total spin $S$ has to obey the triangle rule. The operator under consideration is a vector, thus $|\Delta S|\leq1$ and a transition where $S=0$ in both initial and final state is forbidden.
As the operator is acting on a single site, when $\Delta S=0$ the $S^z=0$ states have a zero matrix element.
\subsection{Interband matrix elements}
The matrix elements will be expressed in terms of $U_{S}(p,p_1,p_2)$. There are several cases to consider for each of the types of solution listed in Appendix~\ref{sec:bae}, and their respective contributions are shown in Appendix~\ref{sec:inter}. The form for a real solution is
\begin{align}
 \label{e17}
& U_{S}(p,p_1,p_2) \equiv L\mathcal{N}_S(p_1,p_2)e^{-\frac{i}2\delta^{S}_{p_1,p_2}}e^{i\frac{\pi}{2} S} \nn
& \times \left[\frac{\sin\left(\frac{1}2(p-p_1+\delta^S_{p_1,p_2}-\pi S)\right)}{\sin\left(\frac12(p-p_1)\right)}\right. \nn
& + \left.\frac{\sin\left(\frac{1}2(p-p_2-\delta^{S}_{p_1,p_2}-\pi S)\right)}{\sin\left(\frac12(p-p_2)\right)} \right].
\end{align}
We also calculate the perturbative correction to the matrix elements to order $\mathcal{O}\left(\alpha\right)$ in Appendix~\ref{sec:pert}.
The relevant matrix elements are given in Table~\ref{tab1}.
\subsection{Intraband matrix elements}
In the two triplon sector, transitions are possible between most combinations of states listed in Appendix~\ref{sec:bae}. The full list is shown in Appendix~\ref{sec:intra}. The result for transitions between real states is
\begin{align}
\label{eq:intra}
&W_{S^\prime,S}(p^\prime_1,p^\prime_2,p_1,p_2)=\nn
&\qquad L^2\mathcal{N}_S(p_1,p_2)\mathcal{N}_{S^\prime}(p^\prime_1,p^\prime_2) e^{\frac{i}2(\delta^S_{p_1,p_2}-\delta^{S^\prime}_{p^\prime_1,p^\prime_2}+(S^\prime-S)\pi)}\nn
&\times\biggl[\frac{\sin(\frac12(p_1-p^\prime_1-\delta^S_{p_1,p_2}+\delta^{S^\prime}_{p^\prime_1,p^\prime_2}-(S^\prime-S)\pi))}{\sin(\frac12(p_1-p^\prime_1))}\nn
&+\frac{\sin(\frac12(p_1-p^\prime_2-\delta^S_{p_1,p_2}-\delta^{S^\prime}_{p^\prime_1,p^\prime_2}-(S^\prime-S)\pi))}{\sin(\frac12(p_1-p^\prime_2))}\nn
&+\frac{\sin(\frac12(p_2-p^\prime_1+\delta^S_{p_1,p_2}+\delta^{S^\prime}_{p^\prime_1,p^\prime_2}-(S^\prime-S)\pi))}{\sin(\frac12(p_2-p^\prime_1))}\nn
&+\frac{\sin(\frac12(p_2-p^\prime_2+\delta^S_{p_1,p_2}-\delta^{S^\prime}_{p^\prime_1,p^\prime_2}-(S^\prime-S)\pi))}{\sin(\frac12(p_2-p^\prime_2))}\biggr].
\end{align}
In the cases that either of the momenta in the first state equals either of those in the second, the corresponding fraction needs to be replaced by
\be-(L-1)e^{\frac{i}2(\pm\delta^S_{p_1,p_2}\mp\delta^{S^\prime}_{p^\prime_1,p^\prime_2}-(S^\prime-S)\pi)}\nonumber .
\ee

\section{Spectral representation and resummation}

The leading contributions to the low-temperature expansion for the
dynamical susceptibility are given by $G_{0,1}=E_{0,1}+F_{1,0}$.
Using the matrix elements from Table~\ref{tab1}, we find that 
to order $\alpha$ we have
 \begin{align}
 \label{es4}
G_{0,1}=&\frac{(1-\cos{Q_\perp})}{4}(1-\alpha\cos{Q_\parallel}) \nn
& \times \left( \frac{1}{\omega+i\eta-\epsilon_{Q_\parallel}}-\frac{1}{\omega+i\eta+\epsilon_{Q_\parallel}}\right).
\end{align}
These give rise to a delta function peak located at the one-triplon
excitation energy. The intraband term $G_{1,1}$ is given by
\begin{equation}
\label{es5}
 G_{1,1}= \frac{(1+\cos{Q_\perp})}{2L} \sum_{p}\frac{e^{-\beta \epsilon_p}-e^{-\beta \epsilon_{Q_\parallel+p}}}{\omega+i\eta+\epsilon_{p}-\epsilon_{Q_\parallel+p}}.
\end{equation}
Similarly, we find the interband terms
\begin{align}
\label{es6}
G_{1,2}=&\frac{(1-\cos{Q_\perp})}{4L^2}
\sum_{p_1>p_2} (e^{-\beta \epsilon_{Q_\parallel+p_1+p_2}}\nn
& \times \left(\frac{1}{\omega+i\eta+\epsilon_{Q_\parallel+p_1+p_2}-\epsilon_{p_1}-\epsilon_{p_2}}\right.\nn
&-\left.\frac{1}{\omega+i\eta+\epsilon_{p_1}+\epsilon_{p_2}-\epsilon_{Q_\parallel+p_1+p_2}}\right) \nn
& \times \sum_S\frac{2S+1}{3}(|U_S^2-\alpha U_S V_S|).
\end{align}

The sum over $p_1,p_2$ is taken over all momenta that satisfy
the boundary conditions \eqref{e12}, and these momenta depend
on $S$. The leading term in $G_{1,2}$ scales with $L$, but
cancels against the ``disconnected" contribution $Z_1G_{0,1}$. The
low-temperature expansion of the dynamical susceptibility now takes
the form 
\be
 \chi^{zz}(\omega, \mathbf{Q})\approx\mathcal{C}_0+\mathcal{C}_1+\mathcal{C}_2,
\label{lowTchi}
\ee
where
\begin{align}
\mathcal{C}_0(\omega,{\bf Q})&\approx G_{0,1},\nn
\mathcal{C}_1(\omega,{\bf Q})&\approx G_{1,0}+G_{1,1}
+\Bigl(G_{1,2}-Z_1G_{0,1}\Bigr),\nn
\mathcal{C}_2(\omega,{\bf Q})&\approx G_{2,2}-Z_1G_{1,1}.
\end{align}
Here $Z_1=3 \sum_{p}e^{-\beta \epsilon _p}$ is the single particle
contribution to the partition function. We note that in $\mathcal{C}_2$
we only have taken into account the intraband processes. 
We observe the following divergences in $\mathcal{C}_n$:
\begin{equation}
\mathcal{C}_n(\omega,Q_\parallel)\propto
\begin{cases}
\left(\frac{1}{(\omega+i\eta)^2-\eps^2(Q_\parallel)}\right)^{1+n}
& \omega^2\approx \epsilon^2_{Q_\parallel},\cr
\left(\frac{1}{\varepsilon^2(Q_\parallel)-(\omega+i\eta)^2}\right)^{n-1/2}
& \omega^2\approx \varepsilon^2_{Q_\parallel},\cr
\end{cases}
\label{singularities}
\end{equation}
where we have defined 
\be
\varepsilon(k)=2J_\parallel\sin(Q_\parallel/2).
\label{veps}
\ee
The first (second) kind of singularity is seen to be present in $\mathcal{C}_n$
for $n=0,1$ ($n=1,2$).
We expect (\ref{singularities}) to hold for $n\geq 2$ as well.
Following the procedure set out in Section~\ref{sec:div}, we define
\begin{align}
\chi^{zz}_{\mathrm{sing},2}&\approx G_{1,1}+(G_{2,2}-Z_1G_{1,1}),\nn
\chi^{zz}_{\mathrm{sing},1}&\approx
G_{0,1}+G_{1,0}+(G_{1,2}-Z_1G_{0,1}).
\end{align}
The leading orders in the low-temperature expansions of the
self-energies then take the form 
\begin{align}
\Sigma_1(\omega, \mathbf{Q}) &= G_{0,1}^{-2}(\omega,
\mathbf{Q})\left[G_{1,2}(\omega, \mathbf{Q})-Z_1G_{0,1}(\omega,
  \mathbf{Q})\right],\nn
\Sigma_2(\omega, \mathbf{Q}) &= G_{1,1}^{-2}(\omega,
\mathbf{Q})\left[G_{2,2}(\omega, \mathbf{Q})-Z_1G_{1,1}(\omega,
  \mathbf{Q})\right].\nn
\end{align}
Our approximate result for the DSF is then
\begin{align}
S^{zz}(\omega,\mathbf{Q})=&
-\lim_{\eta\rightarrow0}\frac1\pi\frac1{1-e^{-\beta\omega}}\nn
&\times\Im\Biggl[
\frac{G_{1,1}(\omega,\mathbf{Q})}
{1-G_{1,1}(\omega,\mathbf{Q})\Sigma_2(\omega,\mathbf{Q})}\nn
&\qquad+ \frac{G_{0,1}(\omega,\mathbf{Q})}
{1-G_{0,1}(\omega,\mathbf{Q})\Sigma_1(\omega,\mathbf{Q})}\Biggr]. 
\label{szz}
\end{align}
\section{Results and discussion}
\label{sec:conc}

In order to present explicit results, we choose $\alpha=0.1$ and
perform numerical calculations on a system of $L=1000$
dimers. Doubling the number did not change the results
significantly. The limit $\eta \rightarrow 0$ is approximated by
choosing a value larger than the spacing of the momentum values due to
finite size, which is of order $\mathcal{O}(\frac{4\pi}{L}J_\parallel)$,
but small compared to the thermal broadening $J_\parallel e^{-\beta
  J_\perp}$, so that the shape of the response is not changed
significantly. One problem we encounter is that to the order in
$J_\parallel/J_\perp$ we are working in, the bound state contributions
to $\mathcal{C}_1$ give rise to sharp peaks for kinematic reasons. These
features will be suppressed once higher orders in perturbation theory
are taken into account, even if we do not sum higher order terms
in the low-temperature expansion (which would lead to a further
broadening). Given that the sharp bound state peaks are an artifact of
the order in perturbation theory considered, we choose to
suppress them in the various plots by specifying a sufficiently large
broadening $\eta=0.01$. This also facilitates comparison to the ED
results. The choice of $Q_\perp$ affects the mixing between the
intraband ($\propto \cos^2 \frac12 Q_\perp$) and interband ($\propto
\sin^2 \frac12 Q_\perp$) responses. Hence plots are given for
$Q_\perp=\pi/2$, where both types of transition are allowed with equal
weight. 
\subsection{Broadening of the triplon line}
\begin{figure}[tbp]
\includegraphics*[width=0.45\textwidth]{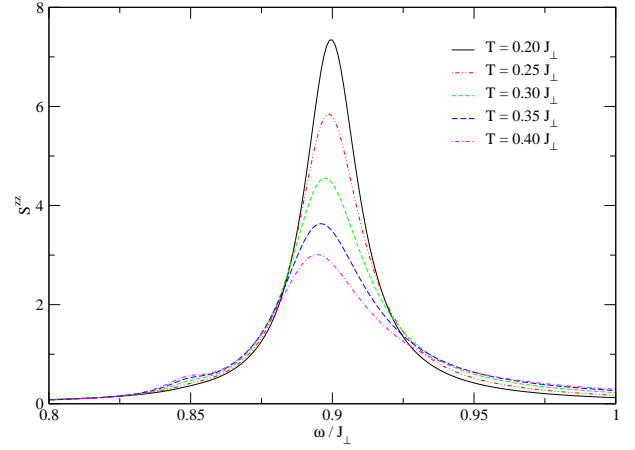}
\caption{The interband transition for $(Q_\parallel,Q_\perp)=(\pi,\pi/2)$
and $L=1000$ sites. The asymmetry grows as $T$ increases.}  
\label{temperature}
\end{figure}
\begin{figure}[tbp]
\includegraphics*[width=0.45\textwidth]{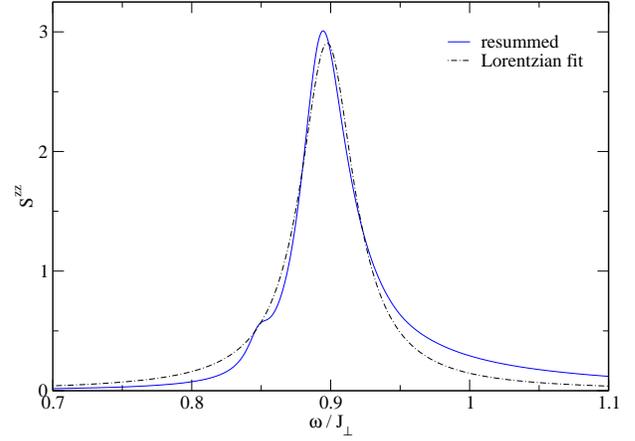}
\caption{The resummed interband transition lineshape for
$T=0.4J_\perp$, $Q_\perp=\pi/2$, $Q_\parallel=\pi$, $\eta=0.01$ and $L=1000$ together with a Lorentzian best fit demonstrating the asymmetric lineshape.}  
\label{fig:lor}
\end{figure}
\begin{figure}[tbp]
\includegraphics*[width=0.45\textwidth]{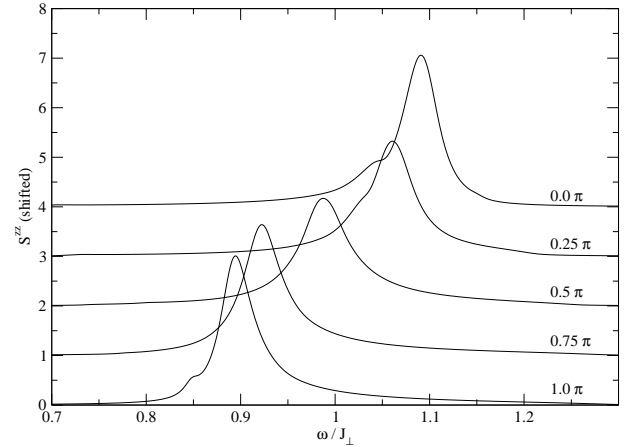}
\caption{Dependence of the interband transition at $T=0.5J_\perp$
on $Q_\parallel$. $Q_\perp$ is fixed at $\pi/2$ and $L=1000$.
The graphs are offset by $n$ for $Q_\parallel=n\pi/4$.}  
\label{fig:position}
\end{figure}
\begin{figure}[tbp]
\includegraphics*[width=0.45\textwidth]{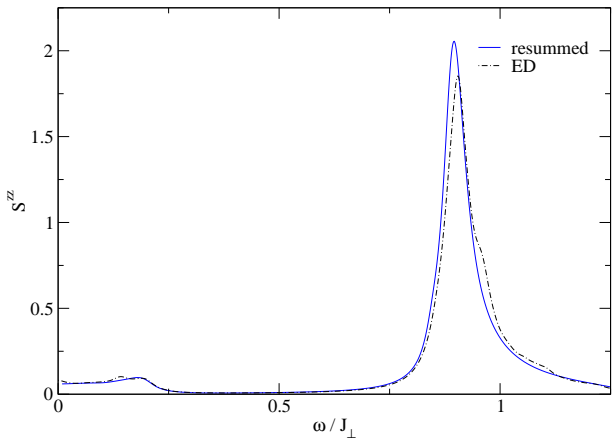}
\caption{A comparison of the resummed spectral function for
$T=0.4J_\perp$, $Q_\perp=\pi/2$, $Q_\parallel=\pi$, $\eta=0.02$ and $L=1000$ to the ED result for a $L=8$ system.}  
\label{fig:ED}
\end{figure}

We first consider the temperature evolution of the triplon line. At
$T=0$ the DSF features a delta function line following
the triplon dispersion.
In Fig.~\ref{temperature}, we plot $S^{zz}(\omega,{\bf Q})$ as a
function of frequency for wave vector ${\bf Q}=(\pi,\pi/2)$ and
temperatures in the range $0.2J_\perp\leq T\leq 0.4 J_\perp$. We see
that the line broadens \emph{asymmetrically} in energy as the
temperature increases. On the other hand, at sufficiently low
temperatures we expect the lineshape to be well approximated by a
Lorentzian \cite{damle}. In Fig.~\ref{fig:lor}, we show a comparison of
the actual result to a Lorentzian fit
\be
S_\mathrm{Lor}(\omega,\mathbf{Q})=
A(\mathbf{Q})
\frac{1/\tau_\phi}{(\omega-\epsilon(Q_\parallel))^2+1/\tau_\phi^2}.
\label{Lorentzian}
\ee
Fig.~\ref{fig:position} shows the dependence of the asymmetry on
$Q_\parallel$. The falloff is slower towards the 
centre of the dispersion.
In order to establish the temperature range in which our
low-temperature expansion provides accurate results, we compare
(\ref{szz}) to numerical results obtained by a direct diagonalization
of the Hamiltonian for short chains. To obtain a
continuous curve for the DSF, we convolve the numerical
results with a Lorentzian of width~$\eta=0.02$. Fig.~\ref{fig:ED} 
shows such a comparison for $T=0.4J_\perp$, $Q_\perp=\pi/2$,
$Q_\parallel=\pi$ and $L=1000$. We see that there is good agreement
between the two methods.
\subsection{Finite temperature resonance at low frequencies}
\begin{figure}[tbp]
\includegraphics*[width=0.45\textwidth]{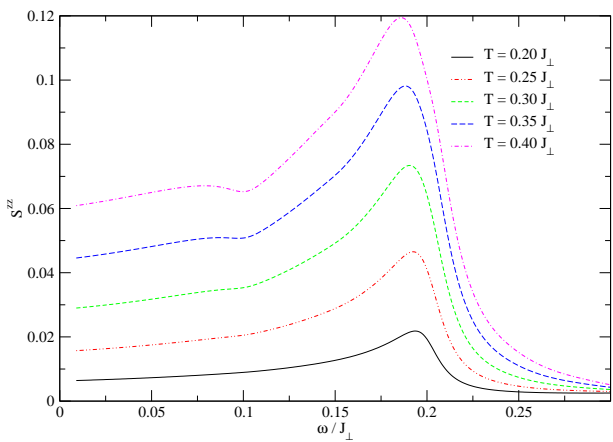}
\caption{The intraband transition at a series of temperatures.
$Q_\perp=\pi/2$, $Q_\parallel=\pi$, $\eta=0.01$ and $L=1000$.}  
\label{fig:intra}
\end{figure}
As in the state of thermal equilibrium there is a finite density of
triplons, incident neutrons can scatter off them with energy transfers
small compared to the gap. Accordingly at finite temperatures there is
a spin response at energies $\omega \sim 0$. To leading contribution
to this ``intraband response" is
\begin{align}
&-\frac{1}{\pi}\frac{1}{1-e^{-\beta \omega}} \Im G_{1,1}=
\frac{1+\cos(Q_\perp)}{2\pi} 
\frac{e^{-\beta(J-\omega/2)}}{\sqrt{\varepsilon^2(Q_\parallel)-\omega^2}}
\nn
&\times\cosh\left(\frac{\beta\cot(Q_\parallel/2)}{2}\sqrt{\varepsilon^2(Q_\parallel)-\omega^2}\right)\theta\left(\varepsilon^2(Q_\parallel)-\omega^2\right),
\end{align}
where $\varepsilon(Q_\parallel)$ is given by (\ref{veps}).
This contribution contains square root singularities for
$\omega\to\pm\varepsilon(Q_\parallel)$, which get smoothened once we resum
terms following Section~\ref{sec:div2}. In Fig.~\ref{fig:intra} we
plot the DSF at low frequencies for several
temperatures in the range $0.2J_\perp\leq T\leq 0.4J_\perp$. We see
that the integrated intensity increases with temperature, while a
strong peak at $\omega\approx\varepsilon(Q_\parallel)$ remains. This
is very similar to what happens in the spin-$\frac12$ Heisenberg-Ising chain
\cite{JGE}, where this feature was first predicted by Villain
\cite{villain}. 
\subsection{Summary}
In this work we have determined the low temperature dynamical
structure factor of the two-leg spin-$\frac12$ Heisenberg ladder in the
limit where the leg coupling is weak compared to the rung exchange.
We have shown that the sharp delta-function line following the triplon
dispersion at $T=0$ gets broadened in an asymmetric way at $T>0$.
The dominant processes at low $T$ involve scattering from one-triplon
to two-triplon states in the presence of a ``thermal background'', as
described in Section~\ref{sec:lowTex}.
We have also determined the temperature activated contribution to the
DSF at low frequencies. Here the dominant processes at
low $T$ involve scattering between different two-triplon states in the
presence of a ``thermal background''.
Our analysis is based on the method developed in
Ref.~[\onlinecite{JEK08}] for the case of the alternating spin-$\frac12$
Heisenberg chain. We have gone beyond Ref.~[\onlinecite{JEK08}] in two
important aspects. Firstly, we have taken into
account all perturbative corrections to the various matrix elements to
order $\mathcal{O}\left(J_\parallel/J_\perp\right)$. This establishes
that higher order perturbation theory in $J_\parallel/J_\perp$ can be
combined with the low-temperature expansion of Ref.~[\onlinecite{JEK08}].
Secondly, we have included
the order $\mathcal{O}\left(e^{-2\beta J_\perp}\right)$ correction 
$G_{2,2}-Z_1G_{1,1}$ to the intraband contribution. This allows us to
describe the low-frequency temperature induced ``resonance'' in a
significantly larger temperature window and demonstrates the difficulties
encountered when dealing with higher orders in the low-temperature
expansion. It would be interesting to compare our results to experiments
on ladder materials. Perhaps the best candidate is
$\rm (C_5H_{12}N)_2CuBr_4$, which is a highly one-dimensional
two-leg ladder material with $\alpha\approx 0.256$ \cite{RueggA,RueggB,RueggC}.
Experimental studies of the temperature evolution of the DSF
for this material are under way \cite{Ruegg2}.

\begin{acknowledgments}
We are grateful to Andrew James and Christian R\"{u}egg for important
discussions. This work was supported by the EPSRC under grant
EP/D050952/1 and the ESF network INSTANS. 
\end{acknowledgments}

\appendix
\section{Linked-cluster expansion for \texorpdfstring{$J_\parallel=0$}{zero J-parallel}}
For $J_\parallel=0$ we are dealing with an ensemble of uncoupled
dimers. The dynamical susceptibility can then be calcuated by
elementary means in the Matsubara formalism. After analytic
continuation we obtain
\be
\chi^{zz}(\omega>0,\mathbf{Q})=
\frac{J_\perp}{2}\frac{1-e^{-\beta J_\perp}}{1+3e^{-\beta J_\perp}}
\frac{1-\cos(Q_\perp)}{(\omega+i0)^2-J_\perp^2}.
\label{full}
\ee
The temperature dependent factor can be expanded at low temperatures
\be
\frac{1-e^{-\beta J_\perp}}{1+3e^{-\beta J_\perp}}
=1-4e^{-\beta J_\perp}+12e^{-2\beta J_\perp}+\ldots.
\ee
We have calculated the first few terms of the low-temperature
expansion (\ref{lowTex})
by working in a product basis of dimer triplet and singlet states. The
leading contribution is
\be
\mathcal{C}_0=G_{0,1}=
\frac{J_\perp}{2}\frac{1-\cos(Q_\perp)}{(\omega+i0)^2-J_\perp^2},
\ee
which correctly reproduces the $T=0$ limit of (\ref{full}).
The next term is
\be
\mathcal{C}_1=G_{1,0}+(G_{1,2}-Z_1G_{0,1}).
\ee
We find by explicit calculation that
\be
G_{1,2}-Z_1G_{0,1}=-3e^{-\beta J_\perp}G_{0,1}.
\ee
This results in
\be
\mathcal{C}_1=-4e^{-\beta J_\perp}G_{0,1},
\ee
which correctly reproduces the first subleading term in \eqref{full}.
The next term is
\begin{align}
\mathcal{C}_2=&(G_{2,1}-Z_1G_{1,0})\nn
&+(G_{2,3}-Z_1G_{1,2}+(Z_1^2-Z_2)G_{0,1}).
\end{align}
We find that
\begin{align}
G_{2,1}-Z_1G_{1,0}&=-3e^{-\beta J_\perp}G_{0,1},\nn
G_{2,3}-Z_1G_{1,2}+(Z_1^2-Z_2)G_{0,1}&=9e^{-2\beta J_\perp}G_{0,1},
\end{align}
which gives
\be
\mathcal{C}_2=12e^{-\beta J_\perp}G_{0,1}.
\ee
This correctly reproduces the second subleading term in \eqref{full}. 
We note that in the limit $J_\parallel=0$ the low-temperature
expansion~\eqref{lowTex} is well defined and does not suffer from the
kind of ``infrared'' divergences present for $J_\parallel\neq 0$. This
is as expected since the spectral function of the full result~\eqref{full}
features a sharp delta-function line even at $T>0$.

\section{Solutions of the BAE}
\label{sec:bae}
\subsection{Real solutions}
To find the two-triplon momenta allowed by the quantization
condition (\ref{e12}), we follow the approach outlined by
James et al.\ \cite{JEK08}. We choose a suitable branch cut such that the
solutions are enumerated by
\be
\label{ea1}
L p_{1,2}=\mp i\ln(-A^S_{p_1,p_2})+2\pi\left[I_{1,2}+\frac{1+(-1)^S}{4}\right],
\ee
where $I_{1,2}$ are integers used to parametrize the equation. This gives $L(L-1)/2$ possible solutions.
To satisfy $p_1>p_2$ we need $I_1 \geq I_2$. In the case of $I_1 \neq I_2$
this is easily solved numerically, although care must be taken
not to double-count solutions.

One must be careful with those solutions where the phase shift is zero. The momenta are then equal to the single triplon momenta and so the matrix elements can be of order $\mathcal{O}(L)$. These solutions occur only for $S=0$ or $2$. For these states the normalization is
\be
\mathcal{N}_S=\left[L(L-2)\right]^{-\frac12}.
\ee

The procedure above does not identify all real solutions in the $S=0$ sector.
The remaining roots are found following Ref.~[\onlinecite{essxxx}].
For large systems Equation~\eqref{ea1} has solutions where $I_1=I_2$.
Whereas the trivial solution $p_1=p_2$ is forbidden by the Pauli
principle, another solution appears very close to the trivial one.
Due to the proximity of the two zeros, the numerical solution of the
equation is difficult.
One method is to rewrite the BAE as a single equation in $x=p_1-p_2$ and
to then
divide by $x$ to eliminate the trivial zero, after which the root finder
converges reliably on the desired solution.

\subsection{Bound states}
There also exist complex solutions $p_{1,2}=x\pm iy$, where the amplitude decays exponentially
as a function of the separation of triplons, corresponding to
bound states. For these the S-matrix elements are real, and
equation (\ref{e12}) becomes
\be
e^{ixL}e^{-yL}+(-1)^S\frac{2\cos(x)+(2-\frac{S}2 (S+1)) e^{-y}}{2\cos(x)+(2-\frac{S}2 (S+1)) e^y}=0.
\ee
For each $x=n\pi/L$ there may exist a zero, and the number of
solutions scales as $L$. The matrix elements for these roots
require special treatment and are given as previously in
terms of
\begin{align}
\mathcal{N}_S(p_1,p_1^*)&=\left[L(L-1)A^S_{p_1,p_1^*}(-1)^S \right.\nn&+\left.L\frac{e^{-y}-e^y(A^S_{p_1,p_1^*})^2}{2\sinh(y)}\right]^{-\frac12}.
\end{align}

\subsection{Singular solutions (type I)}

At this point we still miss $4$ solutions, which occur at
singularities of the quantization conditions. 
Such a solution was described for the spin-$\frac12$ XXX model in
Ref.~[\onlinecite{essxxx}]. For each $S$ sector
there is a solution at $p_{1,2}=\pi/2 \pm i\infty$, corresponding to a
vanishing S-matrix eigenvalue. By introducing a
twist angle $\phi$ the quantization conditions become 
\begin{align}
A^S_{p_1,p_2}e^{i\phi/2}&=(-1)^Se^{iLp_1} ,\nn
e^{i\phi}&=e^{iL(p_1+p_2)}.
\end{align}
This renders the momenta finite, but they cease to be complex
conjugate to one another. Normalizing the wave function and then
taking the limit $\phi \rightarrow 0$ we obtain
\be
\psi^S_{a,b}=(-1)^b(\delta_{a-1,b}-(-1)^S\delta_{a,L-1}\delta_{b,0}).
\ee
It can be verfied by direct calculation that this gives an eigenstate
of the Hamiltonian. The normalization of the state is
\be
\mathcal{N}_S=L^{-\frac12}.
\ee

\subsection{Singular solutions (type II)}
Finally, there is another singular solution in the $S=0$ sector with
$p_1=p_2=\pi$. This solution gives rise to an eigenstate despite the
fact that the two momenta are the same because the phase shift is
ill-defined. Again the limiting wave function can be calculated by
introducing a twist angle, normalizing the state and then taking the
twist angle to zero. The result for the wavefunction and its normalization is
\begin{align}
  \psi^0_{a,b}&=(-1)^{a+b}\\
  \mathcal{N}_0&=\left(\frac{L(L-1)}2\right)^{-\frac12}
\end{align}

\section{Matrix elements}
\subsection{Interband matrix elements}
\label{sec:inter}
The interband matrix elements for the different types of solution are
as follows: 

\begin{enumerate}
\item \emph{Real solutions with zero phase shift:}
\be
U_S(p,p_1,p_2)=-L \mathcal{N}_S
\left(L\left(\delta_{p_1,p}+\delta_{p_2,p}\right)-2\right).
\ee

\item \emph{Bound states:}
\begin{align}
U_{S}(p,p_1,p_1^*)=&L \mathcal{N}_S(p_1,p_1^*) e^{i\frac{\pi S}{2}}\frac{1}{\cosh(y)-\cos(x-p)} \nn 
&\times\Biggl[(1+A^S_{p_1,p_1^*})\cos(x-p-\frac{\pi S}{2})\nn&\qquad-(e^{-y}+A^S_{p_1,p_1^*}e^y)\cos(\frac{\pi S}{2}) \Biggr].
\end{align}

\item \emph{Singular solutions (type I):}
\be
U_S(p)=2 i L \mathcal{N}_S \sin(p).
\ee

\item \emph{Singular solution (type II):}
\be
U_0(p)=L \mathcal{N}_0.
\ee
\end{enumerate}
\subsection{Intraband matrix elements}
\label{sec:intra}
The intraband matrix elements are as follows for transitions between different types of states in the two triplon sector:

\begin{enumerate}
\item \emph{Real $\rightarrow$ Bound:}
\begin{align}
&W_{S^\prime,S}(p^\prime_1,{p^\prime}^*_1,p_1,p_2)\equiv\nn
&\qquad L^2\mathcal{N}_S(p_1,p_2)\mathcal{N}_{S^\prime}(p^\prime_1,{p^\prime}^*_1) e^{\frac{i}2(S^\prime-S)\pi}e^{\frac{i}2\delta^S_{p_1,p_2}}\nn
&\times\biggl[\frac{(A^{S^\prime}_{p^\prime_1,{p^\prime}^*_1}e^y+e^{-y})\cos((S-S^\prime)\frac{\pi}{2}+\frac12\delta^S_{p_1,p_2})}{\cos(p_1-x)-\cosh(y)}\nn
&-\frac{(A^{S^\prime}_{p^\prime_1,{p^\prime}^*_1}+1)\cos(x-p_1-(S-S^\prime)\frac{\pi}{2}-\frac12\delta^S_{p_1,p_2})}{\cos(p_1-x)-\cosh(y)}\nn
&+\frac{(A^{S^\prime}_{p^\prime_1,{p^\prime}^*_1}e^y+e^{-y})\cos((S-S^\prime)\frac{\pi}{2}-\frac12\delta^S_{p_1,p_2})}{\cos(p_2-x)-\cosh(y)}\nn
&-\frac{(A^{S^\prime}_{p^\prime_1,{p^\prime}^*_1}+1)\cos(x-p_2-(S-S^\prime)\frac{\pi}{2}+\frac12\delta^S_{p_1,p_2})}{\cos(p_2-x)-\cosh(y)}\biggr].
\end{align}

\item \emph{Real $\rightarrow$ Singular (type I):}
\begin{align}
&W_{S^\prime,S}(\frac{\pi}2,\frac{\pi}2,p_1,p_2)\equiv\nn
&\qquad L^2\mathcal{N}_S(p_1,p_2)\mathcal{N}_{S^\prime} e^{\frac{i}2(\delta^S_{p_1,p_2}+(S^\prime-S)\pi)}\nn
&\times2\biggl[\cos\left(p_1-\frac{\delta^S_{p_1,p_2}}2-(S^\prime-S)\frac{\pi}{2}\right)\nn
&\qquad+\cos\left(p_2+\frac{\delta^S_{p_1,p_2}}2-(S^\prime-S)\frac{\pi}{2}\right)\biggr].
\end{align}

\item \emph{Real $\rightarrow$ Singular (type II):}
\begin{align}
&W_{0,1}(\pi,\pi,p_1,p_2)\equiv\nn
&\qquad -iL^2\mathcal{N}_1(p_1,p_2)\mathcal{N}_0 e^{\frac{i}2\delta^S_{p_1,p_2}}\nn
&\times\cos\frac{\delta^S_{p_1,p_2}}2\left[\tan\frac{p_1}{2}+\tan\frac{p_2}{2}\right].
\end{align}

\item \emph{Bound $\rightarrow$ Bound:}
\begin{align}
&W_{S^\prime,S}(p^\prime_1,{p^\prime}^*_1,p_1,p^*_1)\equiv\nn
&\qquad L^2\mathcal{N}_S(p_1,p^*_1)\mathcal{N}_{S^\prime}(p^\prime_1,{p^\prime}^*_1)e^{i(x-x^\prime)}\nn
&\times\biggl[e^{-y-y^\prime}\frac{1-(-1)^{S+S^\prime}A^{S}_{p_1,{p}^*_1}A^{S^\prime}_{p^\prime_1,{p^\prime}^*_1}e^{-i(x-x^\prime)+y+y^\prime}}{1-e^{i(x-x^\prime)-y-y^\prime}}\nn
&+e^{-y+y^\prime}\frac{A^{S^\prime}_{p^\prime_1,{p^\prime}^*_1}-(-1)^{S+S^\prime}A^{S}_{p_1,{p}^*_1}e^{-i(x-x^\prime)+y-y^\prime}}{1-e^{i(x-x^\prime)-y+y^\prime}}\nn
&+e^{y-y^\prime}\frac{A^{S}_{p_1,{p}^*_1}-(-1)^{S+S^\prime}A^{S^\prime}_{p^\prime_1,{p^\prime}^*_1}e^{-i(x-x^\prime)-y+y^\prime}}{1-e^{i(x-x^\prime)+y-y^\prime}}\nn
&+e^{y+y^\prime}\frac{A^{S}_{p_1,{p}^*_1}A^{S^\prime}_{p^\prime_1,{p^\prime}^*_1}-(-1)^{S+S^\prime}e^{-i(x-x^\prime)-y-y^\prime}}{1-e^{i(x-x^\prime)+y+y^\prime}}\biggr].
\end{align}

\item \emph{Bound $\rightarrow$ Singular (type I):}
\begin{align}
&W_{S^\prime,S}(\frac{\pi}2,\frac{\pi}2,p_1,p_2)\equiv\nn
&\qquad L^2\mathcal{N}_S(p_1,{p}^*_1)\mathcal{N}_{S^\prime} e^{\frac{i}2(S^\prime-S)\pi}\nn
&\times2\cos(x-(S^\prime-S)\frac{\pi}2)\left(e^{-y}+A^S_{p_1,{p}^*_1}e^y\right).
\end{align}

\item \emph{Bound $\rightarrow$ Singular (type II):}
\be
W_{1,0}(p_1,p^*_1,\pi,\pi)\equiv iL^2\mathcal{N}_1\mathcal{N}_{0}\frac{\sin(x) \left(1+A^1_{p_1,{p}^*_1}\right)}{\cos(x)+\cosh(y)}.
\ee

\item \emph{Singular (type I) $\rightarrow$ Singular (type I):}
\be
W_{S^\prime,S}(\frac{\pi}2,\frac{\pi}2,\frac{\pi}2,\frac{\pi}2)\equiv2L^2\mathcal{N}_S\mathcal{N}_{S^\prime}\delta_{S,S^\prime}. 
\ee

\item \emph{Singular (type I) $\rightarrow$ Singular (type II):}
\be
W_{0,1}(\pi,\pi,\frac{\pi}2,\frac{\pi}2)\equiv-2L^2\mathcal{N}_1\mathcal{N}_0.
\ee
\end{enumerate}
\subsection{Corrections to the interband matrix elements to first order in \texorpdfstring{$\alpha$}{alpha}}
\label{sec:pert}
We expand the states to first order in $\alpha$ and calculate
the corrections to the matrix elements. Firstly we note that $\mathcal{H}_1$ can
only induce transitions between states where the particle number
differs by at most $2$, since each term in the sum only acts on a pair
of adjacent rungs. Secondly, the Hamiltonian is symmetric under
leg-exchange, while a state $\ket{\gamma_s}$ with $s$ particles picks up a sign 
of $(-1)^{L-s}$. This implies $\langle \gamma_r \vert \mathcal{H}_1 \vert \gamma_s \rangle=0$ if $|r-s|$ is odd. Hence the only contributions to first order are from states with a particle
number different by $2$.

\subsubsection{Ground state corrections}
These are given by 
\begin{align}
\label{e18}
\ket{0}'&=\ket{0}+\sum_{\ket{\gamma_2}} \frac{\bra{\gamma_2} \mathcal{H}_1 \ket{0}}{-2J_\perp}\ket{\gamma_2} +\mathcal{O}(\alpha^2)\nn
&=\ket{0}+\frac{\surd3}4\alpha\sum_{a=0}^{L-1} \phi_{a+1,a}^{0,0}\ket{0}+\mathcal{O}(\alpha^2).
\end{align}

\subsubsection{Single particle state corrections}
There are the following contributions from three-particle states
\begin{align}
\label{e19}
\ket{p,m}^{(1)}&=\sum_{\ket{\gamma_3} } \frac{\bra{\gamma_3} \mathcal{H}_1 \ket{p,m}}{-2J_\perp}\ket{\gamma_3}\nn
&=-\frac{\alpha\surd3}{4\sqrt{L}}\sum_{a=0}^{L-1}\sum_{b\neq a,a-1}e^{ipa}\mathrm{d}_a(m)\phi_{b,b+1}^{0,0}\ket{0}.
\end{align}

\subsubsection{Two-particle state corrections}
In the two-particle sector there are contributions
from four-particle states and from the ground state. The
former do not contribute to any of the matrix elements
used in the subsequent calculation to first order, so
they are not calculated. As the Hamiltonian conserves
$S$ and $m$, only the $\ket{p_1,p_2,S=0,m=0}$ state will have
a correction from the ground state. For real solutions this is given by
\begin{align}
\label{e20}
\ket{p_1,p_2,0,0}^{(1)}&=\frac{\bra{0} \mathcal{H}_1 \ket{p_1,p_2,0,0}}{2J_\perp}\ket{0} \nn
&=-\delta_{p_1+p_2,0} \sqrt{\frac{L}{L-1}} \frac{\surd3}4 \alpha(e^{ip_1}-1)\ket{0}.
\end{align}

Hence the corrections to the matrix elements are:
\begin{enumerate}
\item \emph{Real solutions:}
\begin{align}
\label{e21}
&V_S(p,p_1,p_2) \equiv L\mathcal{N}_S(p_1,p_2)e^{-\frac{i}2\delta^{S}_{p_1,p_2}}e^{i\frac{\pi}2 S} \nn
&\times \left[\frac{\sin\left(\frac{1}2(p-p_1+\delta^S_{p_1,p_2}-\pi S)\right)}{\sin\left(\frac12(p-p_1)\right)}\cos\left(\frac{p_1+2p_2-p}2\right)\right. \nn
&+ \frac{\sin\left(\frac{1}2(p-p_2-\delta^{S}_{p_1,p_2}-\pi S)\right)}{\sin\left(\frac12(p-p_2)\right)}\cos\left(\frac{2p_1+p_2-p}2\right)\nn
&+ 3\delta_{S0}\left( 2\cos\left(\frac{p_1-p_2-\delta^0_{p1,p2}}2\right)\cos\left(\frac{p_1+p_2}2\right)\right.\nn
&\qquad \left. \left. +L\delta_{p_1+p_2,0}\left(\cos\left(\frac{2p_1-\delta^0_{p_1,p_2}}2\right)-1\right)\right)  \right].
\end{align}
\item \emph{Real solutions with zero phase shift:}
\begin{align}
&V_{0,2}(p,p_1,p_2)\equiv L\mathcal{N}_S\nn
&\times\biggl[\left(2\cos(\frac{p-p_1}2)-L\delta_{p_1,p}\right)\cos(\frac{p-p_1}2-p_2)\nn
&\qquad+\left(2\cos(\frac{p-p_2}2)-L\delta_{p_2,p}\right)\cos(\frac{p-p_2}2-p_1)\nn
&\qquad+\delta_{S0} 6 \cos(\frac{p_1+p_2}2)\cos(\frac{p_1-p_2}2)\biggr].
\end{align}
\item \emph{Bound states:}
\begin{align}
&V_{S}(p,p_1,p_1^*)\equiv\frac12 L \mathcal{N}_S(p_1,p_1^*)\nn 
&\qquad\times\biggl[ 3(\delta_{S,0}\delta_{2x,0}L+1)\cos x(e^{-y}+A^S_{p_1,p_1^*}e^y)\nn
&+\frac{(1+A^S_{p_1,p_1^*})\cos(2p-3x)}{\cosh(y)-\cos(x-p)} \nn
&+\frac{(e^{-y}+A^S_{p_1,p_1^*}e^y)(\cos(p-\pi S)-\cos(p-2x))}{\cosh(y)-\cos(x-p)} \nn
&-\frac{(e^{-2y}+A^S_{p_1,p_1^*}e^{2y})\cos(x-\pi S)}{\cosh(y)-\cos(x-p)} \biggr].
\end{align}
\item \emph{Singular solutions (type I):}
\begin{align}
V_{0,2}(p,\frac\pi2,\frac\pi2)&\equiv i L\mathcal{N}_S \sin(2p).
\end{align}
\item \emph{Singular solutions (type II):}
\begin{align}
V_0(p,\pi,\pi)&\equiv L\mathcal{N}_S\left[3+2\sin^2 \frac{p}2\right].
\end{align}
\end{enumerate}
The matrix elements with their corrections can be found
in Table~\ref{tab1}. There is also a non-zero correction
to the matrix elements contributing to $E_{0,2}$ and $F_{0,2}$, but
since this term is zero to leading order the correction to the modulus
squared of the matrix element is only second order.

\bibliographystyle{apsrev4-1}

\end{document}